\documentclass[12pt,preprint]{aastex}
\begin{document}
\title{The Axis-Ratio Distribution of Galaxy Clusters in the SDSS-C4 
Catalog as a New Cosmological Probe}
\author{\sc Jounghun Lee}
\affil{Department of Physics and Astronomy, Seoul National University, 
Seoul 151-742, South Korea}
\email{jounghun@astro.snu.ac.kr}
\begin{abstract} 
We analyze the C4 catalog of galaxy clusters from the Sloan Digital Sky Survey 
(SDSS) to investigate the axis-ratio distribution of the projected two 
dimensional cluster profiles. We consider only those objects in the catalog 
whose virial mass is close to $10^{14}h^{-1}M_{\odot}$, with member galaxies 
within the scale radius $1000$ kpc. The total number of such objects turns 
out to be 336. We also derive a theoretical distribution by incorporating 
the effect of projection onto the sky into the analytic formalism proposed 
recently by Lee, Jing, \& Suto.  The theoretical distribution of the cluster 
axis-ratios is shown to depend on the amplitude of the linear power spectrum 
$(\sigma_8)$ as well as the density parameter $(\Omega_{m})$. Finally, fitting 
the observational data to the analytic distribution with $\Omega_{m}$ and 
$\sigma_{8}$ as two adjustable free parameters, we find the best-fitting 
value of $\sigma_{8}=(1.01\pm 0.09)(\Omega_{m}/0.6)^{(0.07\pm 0.02)
+0.1\Omega_{m}}$.  It  is a new $\sigma_{8}\!\!-\!\!\Omega_{m}$ relation, 
different from the previous one derived from the local abundance of X-ray 
clusters. We expect that the axis-ratio distribution of galaxy clusters, 
if combined with the local abundance of clusters, may put simultaneous 
constraints on $\sigma_{8}$ and $\Omega_{m}$. 
\end{abstract} 
\keywords{cosmology:theory --- large-scale structure of universe}

\section{INTRODUCTION}

The standard theory of structure formation based on the cold dark matter 
(CDM) paradigm explains that the large scale structures of the universe like 
galaxies, groups and clusters of galaxies are originated from the primordial 
fluctuations of the matter density field through gravitational instability.  
In this scenario, the observed properties of the large scale structures 
can be used to extract crucial information on the initial conditions of 
the early universe. 

Among the various properties of the large scale structures, it is the 
abundance of galaxy clusters that has attracted most cosmological attentions 
so far.  Since the clusters are relatively young still in the quasi-linear 
regime, their abundance can be inferred from the linear theory 
\citep[e.g.,][]{pre-sch74}. Furthermore, being rare,  the evolution of galaxy 
clusters depends sensitively on the background cosmology, especially on the 
amplitude of the matter power spectrum on scale of $8 h^{-1}$ Mpc 
($\sigma_{8}$) and the density parameter ($\Omega_{m}$). It has been shown 
that by comparing the linear theory prediction with the cluster abundance 
from X-ray or Sunyaev-Zel'dovich (SZ) effect observations one can constrain 
$\sigma_{8}$ and $\Omega_{m}$ \citep{whi-etal93,bon-mye96,eke-etal96,pen98,
hen00,fan-chi01,pie-etal01,sel02}. For instance, \citet{eke-etal96} found a 
relation of $\sigma_{8}=(0.52\pm0.04)\Omega^{-0.52+0.13\Omega_{m}}_{m}$ 
(for a flat universe with non-zero cosmological constant, $\Lambda$) by 
comparing the Press-Schechter prediction with the local abundance of the 
X-ray clusters complied by \citet{hen-arn91}. 

To break the degeneracy between $\sigma_{8}$ and $\Omega_{m}$, however, 
it is necessary to find another relation obtained from another 
observable property of galaxy clusters. One candidate is the shape of 
galaxy clusters.  It has long  been known observationally as well as 
numerically that in general the shapes of galaxy clusters are not so 
much spherical as elliptical \citep{sch-kin78,bin82,wes-etal89,jin-sut02}.  
Furthermore, it was shown by recent numerical simulations that the 
ellipticity distribution of galaxy clusters depend strongly on the background 
cosmology \citep{jin-sut02,hop-etal05}. This numerical finding implies that 
it may be possible to find a new independent relation between $\Omega_{m}$ 
and $\sigma_{8}$ with the ellipticity distribution of galaxy clusters.

Most of the previous approaches to the cluster ellipticity distribution 
were numerical. In order to use the cluster ellipticity distribution as a 
cosmological probe, however,  it is highly desirable to have an analytic 
model derived from physical principles. Although \citet{bar-etal86} 
made a first analytic attempt in the frame of the density peak formalism, 
their model was applicable only in the asymptotic limit of low ellipticity. 
                      
Very recently, \citet[][hereafter, LJS05]{lee-etal05} derived a new analytic 
expression for the halo axis-ratio distribution by combining the density 
peak formalism and the Zel'dovich approximation. They demonstrated that 
the analytic model produces the characteristic behaviors of the halo 
axis-ratio distribution found in simulations, which gives a hope that it 
may provide a theoretical footing for the use of the cluster ellipticity 
distribution as a cosmological probe.  
Our goal here is to find a new functional form of the 
$\sigma_{8}\!\!-\!\!\Omega_{m}$ relation, independent of the previous one 
derived from the cluster abundance, by comparing the LJS05 analytic 
prediction to the axis-ratio distribution of galaxy clusters determined 
from the recent observational catalog.  

The plan of this paper is as follows: In $\S 2$, we provide a brief 
description of the observational data.  In $\S 3$, we explain how to derive 
an analytic axis-ratio distribution of two dimensional projected profiles 
of galaxy clusters from the LJS05 formalism.   In $\S 4$, we find a new 
functional form of relation $\sigma_{8}\!\!-\!\!\Omega_{m}$ relation 
through fitting the model to the observational data. In $\S 5$, we discuss 
the success and the limitation of our work, and draw a final conclusion. 

\section{DATA FROM SDSS-C4 CATALOG}

The C4 catalog contains $748$ objects \citep{mil-etal05} identified 
in the Second Data Release (DR2) of the Sloan Digital Sky Survey 
\citep[SDSS,][]{str-etal02}. The sky coverage and the redshift range of the 
catalog are $\sim 2600$ deg$^{2}$ and $0.02 \le z \le 0.17$, respectively.  
The majority of the objects in this catalog are clusters of galaxies with 
the  mean membership of $36$ galaxies. In the catalog, however, are also 
included smaller groups with the membership of at most $10$ galaxies as well 
as larger structures with the membership of over $100$ {\it clusters}.
The mass range of the catalog is thus quite broad from 
$10^{12}h^{-1}M_{\odot}$to $10^{16}h^{-1}M_{\odot}$ ($h$: the dimensionless 
Hubble parameter).

The SDSS-C4 catalog provides various spectroscopic properties of galaxy 
clusters, among which the following informations are used for our study: 
the axis-ratio of a projected two-dimensional image ($q$), the virial 
mass ($M$), and the redshift ($z$).  As below, we briefly describe how 
these quantities were measured. For the detailed descriptions 
of the C4 catalog and the measurement of $M$ and $q$,  
refer to the web-page: \\
http://www.ctio.noao.edu/$\sim$chrism/current/research/C4/dr2/
SDSS\_PARAMETERS.html.

The axis ratio, $q$, is derived from an elliptical fit to a projected two 
dimensional image of a cluster. The elliptical fit was conducted by the 
diagonalization of the covariance of the positions of all galaxies within 
$1000$ kpc with respect to the cluster center. The center of a cluster 
is also measured using all galaxies within $1000$ kpc.

The virial mass, $M$, of each cluster was computed (assuming the isothermality 
and the spherical shape) as
\begin{equation}
M = \frac{3}{G}\sigma^{2}_{1}r_{v}
\end{equation}
where $\sigma_{1}$ is the line-of-sight velocity dispersion and $r_{v}$ is the 
deprojected three dimensional virial radius of a cluster \citep{cal-etal96}. 
The value of $r_{v}$ is related to the {\it ringwise projected harmonic mean 
radius}, $R_{h}$ \citep[see eq. 3 in,][]{cal-etal96}, as $r_{v}=\pi R_{h}/2$ 
\citep{lim-mat60}. The value of $R_{h}$  depends on which scale radius is 
used in selecting the member galaxies. In the catalog, there are five 
different criteria  used to measure the value of $M$: $w=500$, $1000$, $1500$, 
$2000$, and $2500$ kpc.  

In this paper,  consistently, we use the criterion $w=1000$ kpc. Using the 
fixed scale radius $w=1000$ kpc does not mean that the virial radius of a 
cluster is fixed to be $1000$ kpc for all clusters.  
What it means is that the ringwise projected harmonic mean radius, $R_{h}$, 
was determined using those galaxies within the scale radius $w=1000$ kpc 
from the cluster center. 

But, note that using the fixed scale radius of $w=1000$ kpc 
may systematically affect the determination of the cluster mass, $M$.  
Since the catalog contains various objects in the wide mass range, using the 
fixed scale radius of $w=1000$ can cause either underestimate or overestimate 
of the value of $M$, depending on how massive an object is. To reduce this
kind of possible systematic effect, we focus only on the narrow cluster mass 
range of $10^{13.5}h^{-1}M_{\odot} \le M \le 10^{14.5}h^{-1}M_{\odot}$. 
We find total $336$ such clusters in the catalog with the median redshift 
of $z_{m}=0.08$.

\section{PHYSICAL ANALYSIS}

\subsection{Overview of the Analytic Model}

The LJS05 analytic model for the axis-ratio distribution of dark halos 
adopts two classical theories as its background: the Zel'dovich approximation 
\citep{zel70} and the density peak formalism \citep{bar-etal86}. 

By applying the Zel'dovich approximation, LJS05 assumed that the three 
principal axes of a triaxial halo, $\{a,b,c\}$ (with $a \le b \le c$) are 
related to the three eigenvalues $\{\lambda_{1},\lambda_{2},\lambda_{3}\}$ 
(with $\lambda_{1} \ge \lambda_{2} \ge \lambda_{3}$) of the linear deformation 
tensor, $d_{ij} \equiv \partial_{i}\partial_{j}\phi$ where $\phi$ is the 
perturbation potential: 
\begin{equation}
\label{eqn:abc}
a \propto \sqrt{1 - D_{+}\lambda_{1}}, \qquad
b \propto \sqrt{1 - D_{+}\lambda_{2}}, \qquad
c \propto \sqrt{1 - D_{+}\lambda_{3}},
\end{equation}
where $D_{+}=D_{+}(z)$ is the growth rate of the linear density field. 
Here, the sum of the three eigenvalues equals the linear density contrast, 
$\delta \equiv \Delta\rho/\bar{\rho}$ ($\bar{\rho}$: the mean density): 
$\lambda_{1} + \lambda_{2} + \lambda_{3} = \delta$. 

By employing the density peak formalism, LJS05 also assumed that a region 
in the linear density field will condense out a virialized halo if it 
satisfies the following two conditions: 
\begin{eqnarray}
\label{eqn:sph} 
\delta = \delta_{c} = \delta_{c0}D_{+}(0)/D_{+}(z) , \\ 
\label{eqn:tri} 
\lambda_{3} > \lambda_{0} = \lambda_{c0}D_{+}(0)/D_{+}(z),    
\end{eqnarray} 
where $\delta_{c}$ and $\lambda_{c}$ represent the threshold values of  
$\delta$ and $\lambda$, respectively, with 
$\delta_{c0} \equiv \delta_{c}(z=0)$ and $\lambda_{c0} \equiv 
\lambda_{c}(z=0)$. The value of $\delta_{c0}$ can be theoretically evaluated 
with the help of the top-hat spherical collapse model. For instance,  
$\delta_{c0} \approx 1.686$ for a flat universe with 
closure density \citep[e.g.,][]{kit-sut96}.  

As for the value of $\lambda_{c0}$, it is precisely $0$ in the original 
density peak formalism.  Instead of using this theoretical value, however, 
LJS05 used a positive value of $\lambda_c=0.37$ which was determined 
empirically through fitting.  Their claim was that the break-down of the 
linear theory in the highly nonlinear regime is responsible for the deviation 
of $\lambda_{c0}$ from the theoretical value of $0$ in practice. 

Here, we use the theoretical value of $\lambda_{c0}=0$ rather 
than the empirical value of $\lambda_{c0}=0.37$ for the following reason. 
The value of $\lambda_{c0} = 0.37$ was found empirically by comparing the 
analytic model with the numerical result for the concordance cosmology 
only. Remember that our purpose is to use the analytic distribution as a 
probe of cosmology without having any bias. Hence, we set $\lambda_{c0}$ 
at the original theoretical value $0$ in the rest of this paper. 

Defining two real variables $\mu_{1}$ and $\mu_{2}$ as 
\begin{equation}
\mu_{1} \equiv \frac{b}{c},  \qquad \mu_{2} \equiv \frac{a}{c}.
\end{equation}
LJS05 derived analytically the joint probability density distribution 
of $\mu_{1}$ and $\mu_{2}$ which depends on the cluster mass $M$ and the 
formation epoch $z_{f}$:
\begin{eqnarray}
\label{eqn:ratio_dis}
p(\mu_{1}, \mu_{2} ; M, z_{f})
&=& A\frac{3375\sqrt{2}}{\sqrt{10\pi}\sigma^{5}_{M}}
\exp\left[-\frac{5\delta^{2}_{c}}{2\sigma^{2}_{M}} + 
\frac{15\delta_{c}(\lambda_{1}+\lambda_{2})}{2\sigma^{2}_{M}} -
\frac{15(\lambda^{2}_{1}+\lambda_{1}\lambda_{2}+\lambda^{2}_{2})}
{2\sigma^{2}_{M}}\right] \nonumber \\
&&\times(2\lambda_{1}+\lambda_{2}-\delta_{c})(\lambda_{1}-\lambda_{2})
(\lambda_{1}+2\lambda_{2}-\delta_{c}) \nonumber \\
&&\times \Theta\!\left(\frac{1}{D_{f}-\lambda_{1}}\right)
\Theta[\delta_{c}-(\lambda_{1}+\lambda_{2})]
\left|\frac{(\partial\lambda_{1}\partial\lambda_{2})}
{(\partial\mu_{1}\partial\mu_{2})}\right|, 
\end{eqnarray}
where  $D_{f} \equiv D(z_{f})$, $\sigma$ is the standard variation of the 
linear density field, and $A$ is the normalization factor which satisfies
\begin{equation}
A\int p(\mu_{1}, \mu_{2} ; M, z_{f}) d\mu_{1}d\mu_{2} = 1.
\end{equation}

The relation between $\{\mu_{1},\mu_{2}\}$ and $\{\lambda_{1},\lambda_{2}\}$ 
are given as  
\begin{eqnarray}
\label{eqn:lamu1}
\lambda_{1} &=& \frac{1 + (D_{f}\delta_{c}- 2)\mu^{2}_{2} + 
\mu^{2}_{1}}
{D_{f}(\mu^{2}_{1} + \mu^{2}_{2} + 1)},\\
\label{eqn:lamu2} 
\lambda_{2} &=& \frac{1 + (D_{f}\delta_{c}- 2)\mu^{2}_{1} + 
\mu^{2}_{2}}{D_{f}
(\mu^{2}_{1} + \mu^{2}_{2} + 1)},
\end{eqnarray}
and the Jacobian $\left|(\partial\lambda_{1}\partial\lambda_{2})/
(\partial\mu_{1}\partial\mu_{2})\right|$ was found to be 
\begin{equation}
\label{eqn:jac}
\left|\frac{(\partial\lambda_{1}\partial\lambda_{2})}
{(\partial\mu_{1}\partial\mu_{2})} \right| = 
\frac{4(D_{f}\delta_{c} - 3)^2\mu_{1}\mu_{2}}
{D_{f}^2(\mu^{2}_{1}+\mu^{2}_{2}+1)^{3}}.
\end{equation}

Equation (\ref{eqn:ratio_dis}) gives the probability distribution of the 
two axis-ratios of a triaxial galaxy cluster with mass $M$ formed at $z_{f}$.  
In practice, however, what we measure is not the formation epoch, $z_f$ 
but the observation epoch, $z$, i.e., the redshift at which a cluster 
is observed. Hence, LJS05 rederived an analytic expression for 
$p(\mu_{1}, \mu_{2}; M ; z)$ from equation (\ref{eqn:ratio_dis}) as   
\begin{equation}
\label{eqn:ratiodis_z}
p(\mu_{1}, \mu_{2}; M ; z) = \int_{z}^{\infty}dz_{f}\ 
 \frac{\partial p_{f}(z_{f};2M,z)}{\partial z_{f}}\ 
p(\mu_{1}, \mu_{2} ; 2M ; z_{f}),
\end{equation}
where $\partial p_{f}/\partial z_{f}$ represents the formation epoch 
distribution of galaxy clusters. LJS05 used the fitting formula given by 
\citet{kit-sut96} which is an approximation to the analytic expression 
found by \citet{lac-col94}.

\subsection{Projection onto the Sky}

Although equations (\ref{eqn:ratio_dis})-(\ref{eqn:ratiodis_z}) provides an 
analytic expression for the axis-ratio distribution of the {\it three 
dimensional} shape of a triaxial cluster with mass $M$ observed at $z$,  
it is hard to compare it directly with the observational data. 
In observation, as we noted in $\S 2$, what is measured is only a {\it two 
dimensional} projected profile of a galaxy cluster. Therefore, for a 
proper comparison with observational data, it is necessary to incorporate 
the effect of projection along the line of sight into the analytic model.

To incorporate into equation (\ref{eqn:ratiodis_z}) the projection effect 
onto the plane of the sky, we follow \citet{bin85}:  
Let $(\theta,\phi)$ be the usual polar coordinates of the line-of-sight 
vector in the principal axes of a triaxial cluster, and let $q$ be the 
axis-ratio of a two dimensional cluster profile projected along the line 
of sight direction. Then, the probability density distribution of $q$ 
can be obtained by performing an integration of equation 
(\ref{eqn:ratio_dis}) over $\mu_{1},\mu_{2},\theta,\phi$ as 
\begin{equation}
\label{eqn:2d_ratiodis}
p(q;M,z) = \frac{1}{4\pi}\int_{0}^{2\pi}d\phi\!\int_{0}^{\pi}\sin\theta 
d\theta\!\int_{\mu_{2}}^{1}d\mu_{1}\!\int_{0}^{1}d\mu_{2}~
\delta_{D}[q-q^{\prime}
(\theta,\phi,\mu_{1},\mu_{2})]~p(\mu_{1},\mu_{2};M,z),
\end{equation}
where
\begin{equation}
\label{eqn:q_def}
q^{\prime}(\theta,\phi,\mu_{1},\mu_{2}) = 
\left\{\frac{U+V - \left[(U-V)^{2}+W^{2}\right]^{1/2}}
{U+V + \left[(U-V)^{2}+W^{2}\right]^{1/2}}\right\}^{1/2},
\end{equation}
and 
\begin{eqnarray}
\label{eqn:a_def}
U &\equiv& \frac{\cos^{2}\theta}{\mu_{2}^{2}}\left(\sin^{2}\theta + 
\frac{\cos^{2}\phi}{\mu_{1}^{2}}\right)+\frac{\sin^{2}\theta}{\mu_{1}^{2}},\\
\label{eqn:b_def}
W &\equiv& \cos\theta\sin2\phi\left(1-\frac{1}{\mu_{1}^{2}}\right)
\frac{1}{\mu_{2}^{2}},\\
\label{eqn:c_def}
V &\equiv& \left(\frac{\sin^{2}\phi}{\mu_{1}^{2}} + \cos^{2}\phi\right)
\frac{1}{\mu_{2}^{2}}.
\end{eqnarray}
Through equations (\ref{eqn:ratio_dis})-(\ref{eqn:c_def}), 
one can evaluate analytically the probability density that the projected 
profile of a galaxy cluster of mass $M$ is observed to have an axis-ratio 
of $q$ at redshift $z$. This theoretical axis-ratio distribution varies 
with the background cosmology since it depends on the initial power spectrum, 
$P(k)$. In this paper, we adopt the following approximation formula for 
$P(k)$ \citep{bar-etal86}:
\begin{equation}
P(k) \propto k\left[\frac{\ln(1+2.34q)}{2.34q}\right]^{2}
[1 + 3.89q + (16.1q)^{2} + (5.46)^{3} + (6.71q)^{4}]^{-1/2},
\end{equation}
where $q \equiv k/\Gamma h^{-1}$ Mpc. Here, the shape factor, $\Gamma$,  
is related to $\Omega_{m}$, $h$, and the baryon density parameter, 
$\Omega_{b}$ \citep{pea-dod94,sug95} as
\begin{equation}
\label{eqn:gamma}
\Gamma = \Omega_{m}h\left(\frac{T_{0}}{2.7 K}\right)
\exp\left[-\Omega_{b}\left(1 + \sqrt{2h}\Omega_{m}^{-1}\right)\right],
\end{equation}
where $T_{0}$ is the temperature of the cosmic microwave background radiation.

\begin{figure}
\begin{center}
\plotone{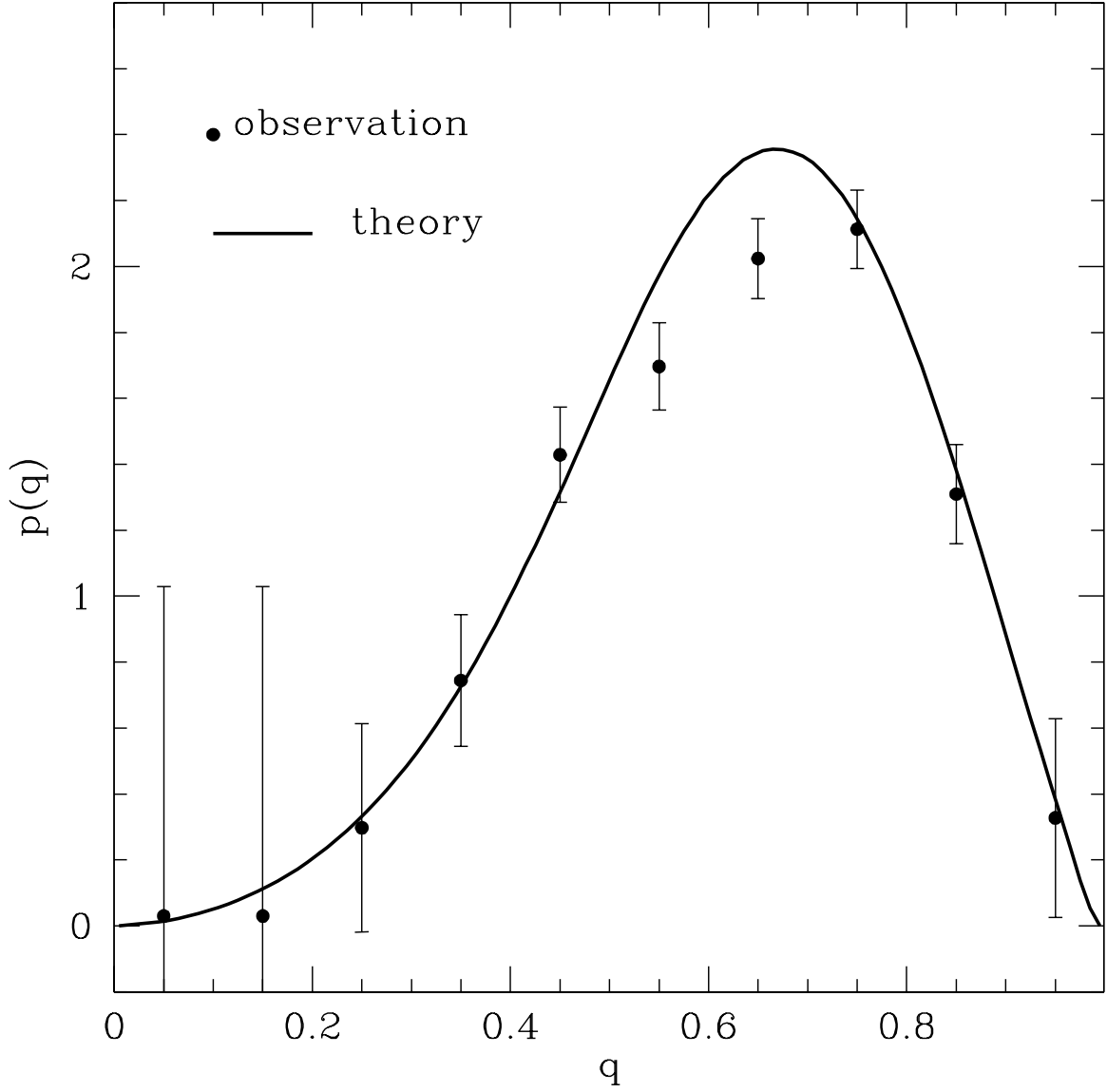}
\caption{The analytic axis-ratio distribution of the two dimensional 
projected profiles of galaxy clusters (solid line) and the observational 
data points arising from the SDSS-C4 catalog (solid dots). 
The error bars are Poissonian. 
\label{fig:distribution}}
\end{center}
\end{figure}

Fig. \ref{fig:distribution} plots equation (\ref{eqn:2d_ratiodis}) on 
{\it cluster scale}, $M_8$, (see $\S 4$ for a definition of $M_{8}$) 
at present redshift ($z=0$) for four different cosmological models:  
$\Lambda$CDM (solid line); OCDM (dotted); SCDM (dashed);  
$\tau$CDM \citep[][long-dashed]{whi-etal95}. The cosmological parameters 
used to characterize each model are listed in Table 1. 

As one can see, equation (\ref{eqn:2d_ratiodis}) depends sensitively on 
the background cosmology.  More elliptical clusters are predicted by the 
$\Lambda CDM$ and OCDM models with high $\sigma_{8}$ and low $\Omega_{m}$ 
than by the SCDM and $\tau$CDM models with low $\sigma_{8}$ and high 
$\Omega_{m}$.   It can be understood by the following logic: the SCDM and 
$\tau$CDM models predict more small scale powers, and thus should result in 
rounder shapes of galaxy clusters in the end.  

We compare equation (\ref{eqn:2d_ratiodis}) with the observational 
distribution from the 336 SDSS-C4 clusters, assuming the $\Lambda$CDM. 
Figure \ref{fig:distribution} plots together the analytic (solid line) 
and the observational (dots) axis-ratio results for comparison. 
The errors of the observational points are Poissonian. 
The values of $M$ and $z$ in the analytic axis-ratio distribution are set 
to be $M_{8}$ (see \S 4.1) and $z_{med} = 0.08$, to be consistent with the 
observational data.  As one can see, the overall agreement between the two 
is quite good.  

\section{APPLICATION OF THEORY TO OBSERVATIONS}

\subsection{Characteristic Cluster Mass}

The analytic axis-ratio distribution (eq.[\ref{eqn:2d_ratiodis}]) depends 
on the cluster mass $M$ and redshift $z$.  Therefore, to compare it with the 
observational data,  one has to first specify the values of $M$ and $z$. 
As for $z$, to be consistent with the observational data, we set it at the 
median redshift of the selected 336 clusters, $z_{m}=0.08$, given that 
the redshift range of the selected clusters is quite narrow. 

As for $M$, there is a subtle point that has to be taken into consideration. 
The virial mass, $M$, of each cluster in the C4 catalog was computed under 
the assumption of a {\it concordance} cosmology with $\Omega_{m}=0.3$, 
$\sigma_{8}=0.9$, $h=0.7$. In other words, the cluster mass $M$ given in 
the catalog is biased toward the concordance cosmology, which has to be 
avoided in the parameter estimation. Thus, we set $M$ at the 
{\it characteristic mass scale}, $M_{8}$, defined as 
\begin{equation}
\label{eqn:m8}
M_{8} = \frac{4\pi}{3}\bar{\rho}R^{3}_{8},
\end{equation}
where $R_{8} = 8h^{-1}$Mpc is the top-hat spherical radius on which scale 
the rms density fluctuation is {\it observed} to be very close to unity, 
and $\bar{\rho}=2.78\cdot 10^{11}\Omega_{m}h^{2}M_{\odot}$Mpc$^{-3}$. 
Since the galaxy clusters are the largest collapsed objects in the universe, 
the characteristic mass scale $M_{8}$ is believed to represent the typical 
mass of galaxy clusters.  Note that this characteristic cluster mass, $M_{8}$, 
by its definition (eq.[\ref{eqn:m8}]), depends on the value of $\Omega_{m}$.  
Therefore, putting $M=M_{8}$ into equation (\ref{eqn:2d_ratiodis}), will 
optimize the parameter estimation without bearing any bias toward a certain 
cosmology. 

\subsection{Constraint on $\Omega_{m}$ and $\sigma_{8}$ for a Flat Universe }

The analytic distribution (\ref{eqn:2d_ratiodis}) depends not only on 
$\Omega_{m}$ and $\sigma_{8}$ but also on $\Omega_{\Lambda}$, $h$, and 
$\Gamma$ as well. Here, we focus  mainly on constraining $\Omega_{m}$ and 
$\sigma_{8}$, assuming that the other parameters are already known as priors. 
We use the following priors from the CMB observation 
\citep[e.g.,][]{lan-etal01}, the HST Key Project \citep{fre-etal01} 
and the big bang nucleosynthesis \citep{oli-etal00}:  
$\Omega_{\rm \Lambda}=1-\Omega_{m}$ (a flat universe); 
$h=0.7$; $\Omega_{b}=0.044$.  Once the values of $h$ and $\Omega_{b}$ are 
given, the shape factor, $\Gamma$, will depend only on the value of 
$\Omega_{m}$ (see eq.[\ref{eqn:gamma}]). 

Now that all parameters except for $\Omega_{m}$ and $\sigma_{8}$ in 
equation (\ref{eqn:2d_ratiodis}) are prescribed, we can fit the 
observational data points obtained in $\S 2$ to equation 
(\ref{eqn:2d_ratiodis}) with adjusting $\Omega_{m}$ and $\sigma_{8}$ 
in ranges of $(0,1]$ and $(0,2]$, respectively.  
We pick as the best-fit values of $\Omega_{m}$ and $\sigma_{8}$ those that 
minimize $\chi^{2}$ : 
\begin{equation}
\chi^{2} = \sum_{i=1}^{N_{p}}\frac{[p_{ob}(q_i)-p(q_i;M_{8},z_{m})]^{2}}
{\sigma^{2}_{ob}(q_i)},
\end{equation}
where $N_{p}$ is the total number of observational data points, 
$\{q_i,p_{ob}(q_i)\}$, and $\sigma_{ob}(q_i)$ is the observational 
standard variation of $q_i$. 
 
\begin{figure}
\begin{center}
\plotone{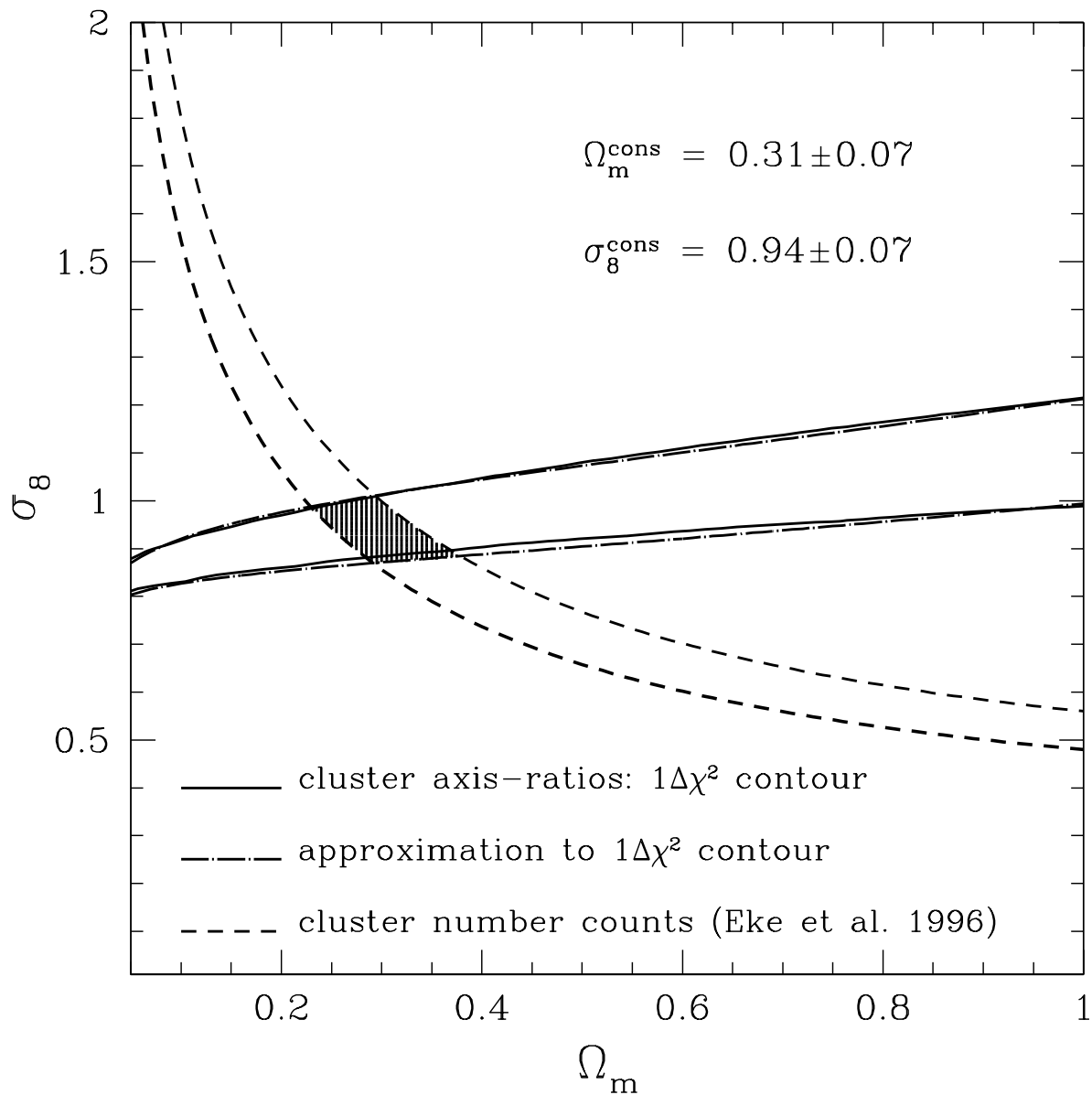}
\caption{The best-fit values of $\sigma_{8}\!\!-\!\!\Omega_{m}$ through 
$\chi^{2}$-statistics. The solid and the dot-dashed lines represent 
the one standard deviation contour and its approximation. The dashed 
lines is the previous relation found from the local abundance of X-ray 
clusters.
\label{fig:contour}}
\end{center}
\end{figure}

The single best-fit values, however, are unreliable given the uncertainty in 
$\chi^{2}$ due to the strong correlation between $\Omega_{m}$ and 
$\sigma_{8}$. A more general best-fit is defined as the full range of the 
outer limit of one standard deviation contour.  
Figure \ref{fig:contour} plots the one standard deviation contour as 
solid line.  We find that this contour is well approximated by the relation, 
$\sigma_{8}=(1.01\pm 0.09)(\Omega_{m}/0.6)^{(0.07\pm 0.02)+0.1\Omega_{m}}$. 
This approximation formula is also plotted as dot-dashed line in 
Figure \ref{fig:contour}.  The previous relation of $\sigma_{8} = 
(0.52\pm0.04)\Omega^{-0.52+0.13\Omega_{m}}_{m}$ obtained by \citet{eke-etal96} 
from the cluster abundance is also plotted for comparison as dashed line. 
The shaded area surrounded by the solid and the dashed lines 
in Figure \ref{fig:contour} represents the simultaneously constrained 
best-fit values of $\Omega_{m}$ and $\sigma_{8}$: 
$\Omega_{m}^{cons}=0.31\pm 0.07$ and $\sigma_{8}^{cons}=0.94\pm 0.07$, 
which is in good agreement with the results from the first year {\it WMAP} 
(Wilkinson Microwave Anisotropy Probe) measurements \citep{spe-etal03}. 

\section{DISCUSSION AND CONCLUSIONS}

We have found a new $\sigma_{8}\!\!-\!\!\Omega_{m}$ relation by comparing 
the LJS05 analytic model for the axis-ratio distribution of galaxy clusters 
with the observational data from the SDSS-C4 catalog. 

The success of our final result, however, is subject to a couple of caveats. 
The first caveat comes from our assumption that the principal axes of the 
SDSS-C4 clusters derived from the elliptical fits to the galaxy distributions 
are directly comparable to that of the analytically derived shapes of dark 
halos. This is an obvious simplification of the reality. 

Notwithstanding, the following observational and numerical clues 
should provide a justification for this simplified assumption. 
Many observations revealed that the spatial distribution of cluster galaxies 
and the major axes of their host clusters are arranged in a collinear way
\citep{wes-bla00,pli-bas02,pli-etal03,per-kuh04}.
Recent numerical simulations also demonstrated that the dark halo 
substructures are preferentially located along the major axes of their host 
halos \citep{kan-etal05,lib-etal05,zen-etal05,lee-kan06}. Very recently, 
\citet{lee-kan06} showed by N-body simulations that the triaxial shapes of 
dark matter halos can be reconstructed from the the anisotropic spatial 
distribution of their substructures. Given these empirical findings, one may  
expect that the cluster principal axis derived from the elliptical fit to the 
galaxy distribution is comparable to the analytical one derived from the  
the smooth dark matter distribution.

The second caveat lies in the limited validity of the analytic model. As 
shown by LJS05, the prediction of our analytic model on the cluster 
ellipticity-mass relation disagrees with the numerical finding. 
Our model predicts that the cluster ellipticity decreases with mass, 
while many different numerical simulations \citep{bul02,jin-sut02,kas-evr05,
hop-etal05, bai-ste05} found that the cluster ellipticity actually increases 
with mass. 

In spite of this limitation of the LJS05 analytic model, it is still true 
that the LJS05 tested their model against N-body simulations and showed
that it works quite well, reproducing the characteristic behaviors of the 
numerical result, on a single mass scale (or in a narrow mass range). 
In other words, although the LJS05 model fails in predicting the "change" 
of the axis-ratio distribution in the wide range of halo mass, 
it can still be used to predict the shape of the axis-ratio distribution 
of a dark halo with a given mass. 

As we consider here only those objects in the C4 catalog whose mass lies in 
a narrow range $\sim 10^{14} h^{-1}M_{\odot}$ for the comparison with the 
analytic model, the LJS05 model is expected to be a useful approximation. 
 
Definitely, however, it will be desirable and necessary to refine our 
model more realistically before applying it to a wide range of mass of 
cosmic structures.
Finally, we conclude that it is possible in principle to constrain both  
$\Omega_{m}$ and $\sigma_{8}$ concurrently from the observation of 
galaxy clusters alone, as the cluster shape and abundance distributions 
provide two different $\sigma_{8}\!\!-\!\!\Omega_{m}$ relations. 
The amplitude of the linear power spectrum obtained from the shape 
and abundance distribution of the galaxy clusters is consistent with 
the high value estimated from the recent {\it WMAP} measurements.

\acknowledgments

We thank an anonymous referee for careful reading and many helpful suggestions
that helped us greatly improve the original manuscript. We acknowledge 
stimulating discussions with R. Sheth, M. Im, C.Park, Y.Jing, Y. Choi, 
H. Hwang, J. J. Lee. This work is supported by the research grant No. 
R01-2005-000-10610-0 from the Basic Research Program of the Korea Science 
and Engineering Foundation.


\clearpage
\begin{table}
\caption{The values of the cosmological parameters for each model}
\begin{center}
\begin{tabular}{lccccc}\hline \hline
model & $\Omega_{m}$ & $\Omega_{\Lambda}$ & $\sigma_{8}$ & $h$ & $\Gamma$ 
\\ \hline
$\Lambda$CDM  & 0.3   & 0.7 & 0.9 & 0.7 & 0.168 \\ 
OCDM  & 0.3   & 0 & 0.87 & 0.83 & 0.25 \\
SCDM   & 1 & 0 & 0.52  & 0.5 & 0.5  \\ 
$\tau$CDM   & 1 & 0 & 0.52 & 0.5 & 0.25 \\ \hline
\end{tabular}
\end{center}
\label{table:3d_coef}
\end{table}

\clearpage
\begin{figure}
\begin{center}
\plotone{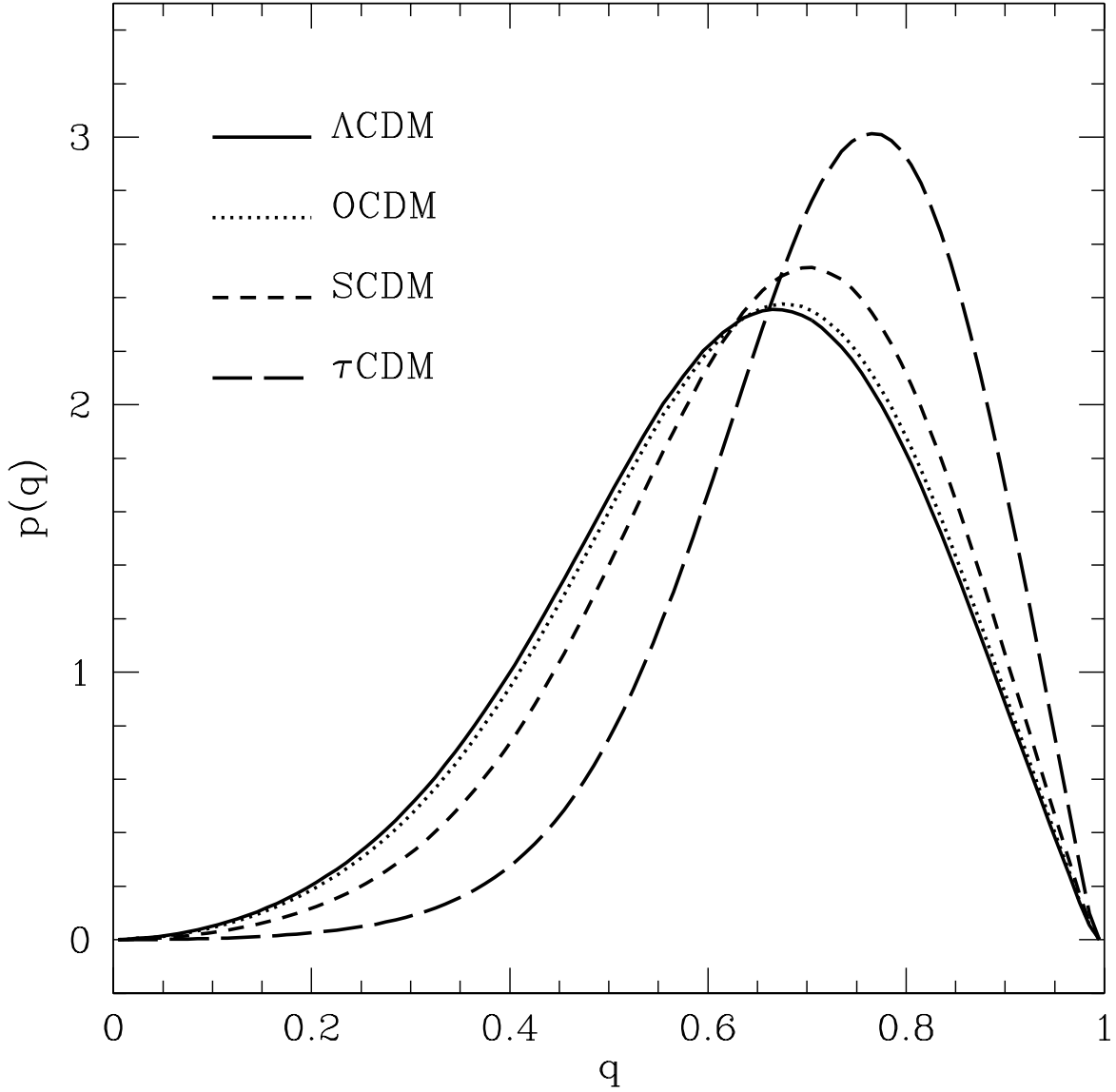}
\caption{The axis-ratio distributions of the two dimensional projected 
profiles of galaxy clusters predicted by the analytic LJS05 model 
(eq.[\ref{eqn:2d_ratiodis}]) at present epoch for four different 
cosmological models: a $\Lambda$CDM model with $\Omega_{m}=0.3$,
$\Omega_{\Lambda}=0.7$,$\sigma_{8}=0.9$,$h=0.7$, $\Gamma=0.168$ (solid line); 
OCDM model with $\Omega_{m}=0.3$, $\Omega_{\Lambda}=0$, $\sigma_{8}=0.87$, 
$h=0.83$, $\Gamma=0.25$ (dotted); SCDM model with $\Omega_{m}=1$, 
$\Omega_{\Lambda}=0$, $\sigma_{8}=0.52$, $h=0.5$, $\Gamma=0.5$ (dashed);
$\tau$CDM model with $\Omega_{m}=1$, $\Omega_{\Lambda}=0$, $\sigma_{8}=0.52$, 
$h=0.5$, $\Gamma =0.25$ (long-dashed).
\label{fig:parameter}}
\end{center}
\end{figure}

\clearpage

\clearpage
\end{document}